\documentclass[sn-nature]{sn-jnl}

\usepackage{times,amsmath,amsfonts,amssymb,latexsym}
\usepackage{graphicx,epsf}
\usepackage{subfigure}
\usepackage{cancel}
\usepackage{color,xcolor}
\usepackage{booktabs}
\usepackage{multirow}
\usepackage{amsthm}
\usepackage{mathrsfs}
\usepackage[title]{appendix}
\usepackage{textcomp}
\usepackage{manyfoot}
\usepackage{algorithm}
\usepackage{algorithmicx}
\usepackage{algpseudocode}
\usepackage{listings}
\usepackage{array}
\usepackage{tabularx}
\usepackage{comment}
\usepackage{lineno}

\begin{document}

\title[Impact of magnetic field direction]
{Impact of magnetic field direction on anti-dot-based superconducting diodes}

\author*[1,2]{\fnm{Eloi Benicio} \sur{de Melo Junior}}\email{eloi.junior@ufv.br}

\author[2]{\fnm{Elia} \sur{Strambini}}

\author[2]{\fnm{Francesco} \sur{Giazotto}}

\author[1]{\fnm{Clodoaldo I. L.} \sur{de Araujo}}

\affil[1]{\orgdiv{Departamento de Física, Laboratório de Spintrônica e Nanomagnetismo},
\orgname{Universidade Federal de Viçosa},
\orgaddress{\city{Viçosa}, \postcode{36570-900},
\state{Minas Gerais}, \country{Brazil}}}

\affil[2]{\orgname{NEST, Istituto Nanoscienze-CNR, Scuola Normale Superiore},
\orgaddress{\city{Pisa}, \postcode{100190},
\country{Italy}}}




\abstract{
The superconducting diode effect (SDE) is a fundamental building block for dissipationless nonreciprocal electronics, yet its microscopic origins in thin films often involve competing mechanisms that remain debated. Here, we demonstrate that the SDE can be engineered in niobium films by patterning macroscopic asymmetric antidots, revealing distinct control mechanisms under in-plane and out-of-plane magnetic fields. We identify two dominant contributions to nonreciprocal transport: edge flux pinning, which governs the low-field and in-plane field regimes via surface-barrier asymmetry, and bulk flux pinning, which drives the high-field response and correlates directly with the geometric asymmetry of the antidots. Supported by time-dependent Ginzburg-Landau simulations and an analytical model, we provide a unified description of these regimes, linking the diode efficiency to the specific pinning landscape. These findings establish a flexible design principle for engineering superconducting diodes with tunable functionality, paving the way for their integration into next-generation quantum and cryogenic circuits.
}
\maketitle

\section{Introduction}\label{sec1}

The simultaneous breaking of time-reversal and inversion symmetries in a superconducting device can give rise to a nonreciprocal response of the supercurrent, known as the supercurrent diode effect (SDE). 
Time reversal is typically broken via magnetic fields, while several designs have been proposed to break the inversion symmetry in either the device geometry, material, or setup \cite{Daido_Intrinsic_2022,Ilic_Theory_2022,Yuan_Supercurrent_2022,He_Phenomenological_2022,Scammel_Theory_2022,fukaya_supercurrent_2025,chirolli_diode_2025,margineda_sign_2023, margineda_back-action_2025,greco_double_2024}.
Indeed, SDE has been reported in a wide variety of systems, including Josephson junctions based on strong spin-orbit semiconductors~\cite{Baumgartner_Supercurrent_2022,superconducting,turini_josephson_2022}, in transition metal dichalcogenides~\cite{supercurrent2}, twisted multilayer graphene~\cite{twist1, twist2}, and topological semimetals~\cite{josephson_semimetal} or driven by asymmetric microwave signals~\cite{borgongino_biharmonic-drive_2025}. A comprehensive overview of these experimental and theoretical findings can be found in recent reviews~\cite{Nadeem2023,shaffer_theories_2025}.

In pristine conventional superconducting films under an out-of-plane magnetic field, the SDE often arises from asymmetry in the vortex-barrier pinning potential with respect to the direction of vortex motion. This asymmetry can be intentionally engineered through patterned antidots~\cite{nanoholes} or may arise intrinsically during fabrication, particularly at the edges of the film~\cite{edgebarriers,Moodera_ubq}. The combination of asymmetric vortex pinning and supercurrent crowding has recently inspired the development of devices specifically designed to exploit vortex-based SDE phenomena~\cite{antola_streamline_2025,castellani_superconducting_2025}, which are further enhanced in narrow geometries~\cite{clem2_8, clem2_9}.

By contrast, the response of superconducting thin films to an in-plane magnetic field differs significantly from the out-of-plane case, as noted in early seminal studies~\cite{Tinkham_Effect_1963,stejic_effect_1994} and reinforced by experimental observations of unconventional enhancement behaviors~\cite{Gardner_Enhancement_2011}. These works laid the foundation for recent investigations of SDE in in-plane fields, with notable examples including Dirac semimetal heterostructures~\cite{Anh_Large_2024}, Nb/V/Ta superlattices~\cite{Ando_Observation_2020,miyasaka_observation_2021}, Pt/V/EuS multilayers~\cite{Moodera_ubq}, and S/N bilayers~\cite{sundaresh_diamagnetic_2023}. In these systems, the SDE is primarily understood through symmetry arguments, specifically magnetochiral anisotropy, consistent with a maximal SDE for fields orthogonal to the supercurrent; however, a comprehensive microscopic picture is often lacking.

Here, we investigate the SDE in Nb films patterned with micrometer-sized antidots. The defects are designed with varying degrees of spatial asymmetry, ranging from fully symmetric circles to drop-like and triangular shapes. An external magnetic field is applied to break the time-reversal symmetry and induce the diode effect. First, we examine the role of the out-of-plane field ($H_z$) and its interplay with different antidot geometries, revealing distinct vortex dynamics across varying field amplitudes. Second, we focus on the influence of in-plane fields ($H_{\text{in}}$) and analyze the dependence of the SDE on both the antidot shape and the field orientation in a wide range of strengths. Finally, we explore the combined effect of $H_z$ and $H_{\text{in}}$ to elucidate their respective contributions and identify potential synergies to control SDE.

\section*{Results}\label{sec2}
\begin{figure}[H]
    \centering
    \includegraphics[width=\linewidth]{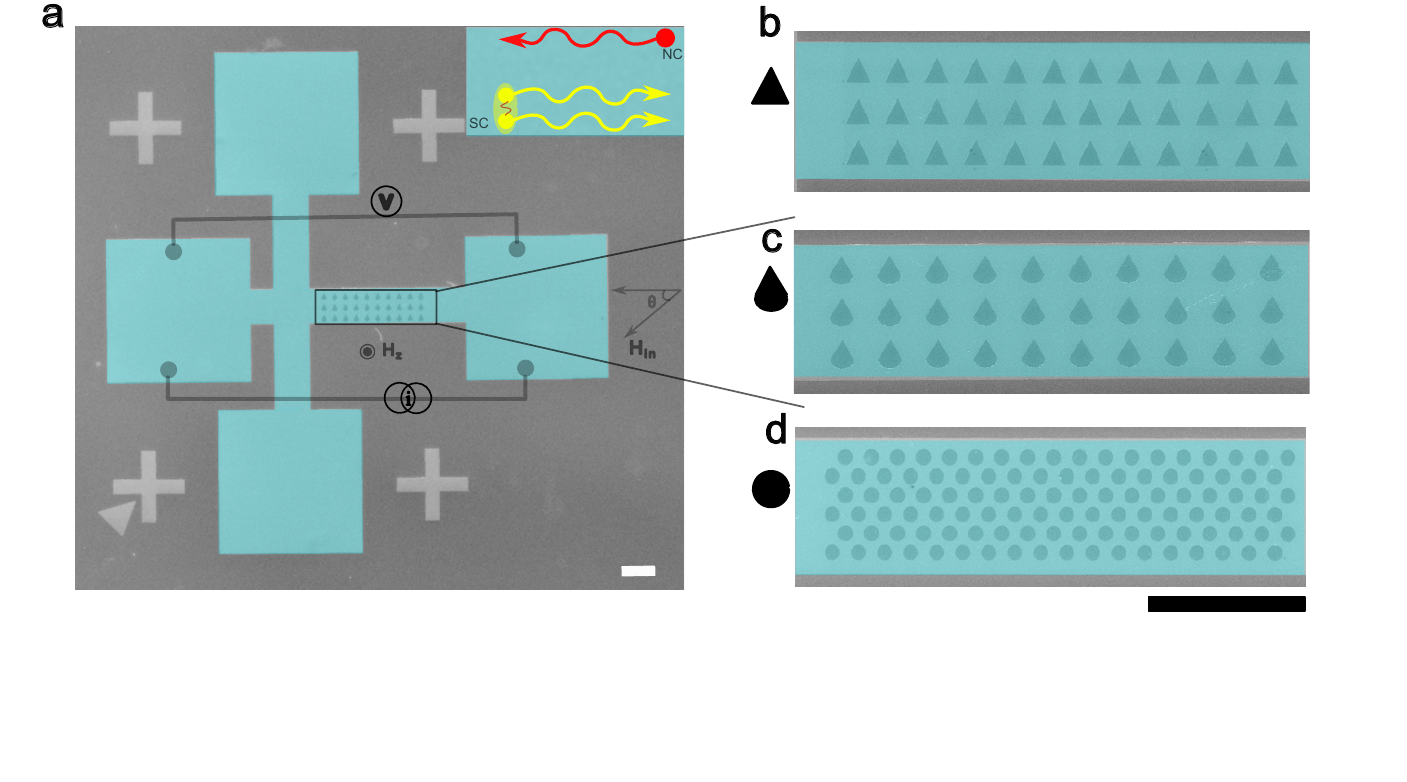}
    \caption{\textbf{Colored SEM image of one of the devices and experimental setup.}  
(a) Measurement scheme and sample overview, consisting of a Nb Hall-bar thin film with a thickness of 200~nm and a patterned region of length 200~$\mu$m.  
The patterned defects exhibit increasing geometrical asymmetry, as shown in the zoomed panels:  
(b) circular holes, (c) drop-shaped antidots, and (d) triangular antidots.  
The white and black scale bars correspond to 50~$\mu$m.  
The inset in panel (a) illustrates the supercurrent diode effect (SDE), showing dissipationless supercurrent in one direction and dissipative current in the opposite direction.}
    \label{samples}
\end{figure}

Our devices consist of thin Nb films patterned with antidots to control vortex dynamics by introducing different degrees of geometrical asymmetry, thereby enhancing the SDE, along with a pristine reference sample. 
The devices are fabricated from Nb stripes with thickness $d = 200$~nm and width $W = 50~\mu$m, patterned in a cross shape as shown in Fig.~\ref{samples}a (fabrication details are provided in the Methods section). Arrays of antidots with different geometries are implemented on the long arm of the cross, ranging from symmetric circles to drop-shaped and asymmetric triangles, as illustrated in the SEM images in Fig.~\ref{samples}b-c. The $VI$ characteristics of the devices were measured using a four-wire configuration to determine the switching critical currents under positive and negative bias, denoted $I_c^+$ and $I_c^-$, respectively. These values were then correlated with the degree of antidot asymmetry under an applied magnetic field.

\subsection*{Out-of-plane magnetic field response.}\label{subsecOUT}

\begin{figure}[H]
    \centering
    \includegraphics[width=\linewidth]{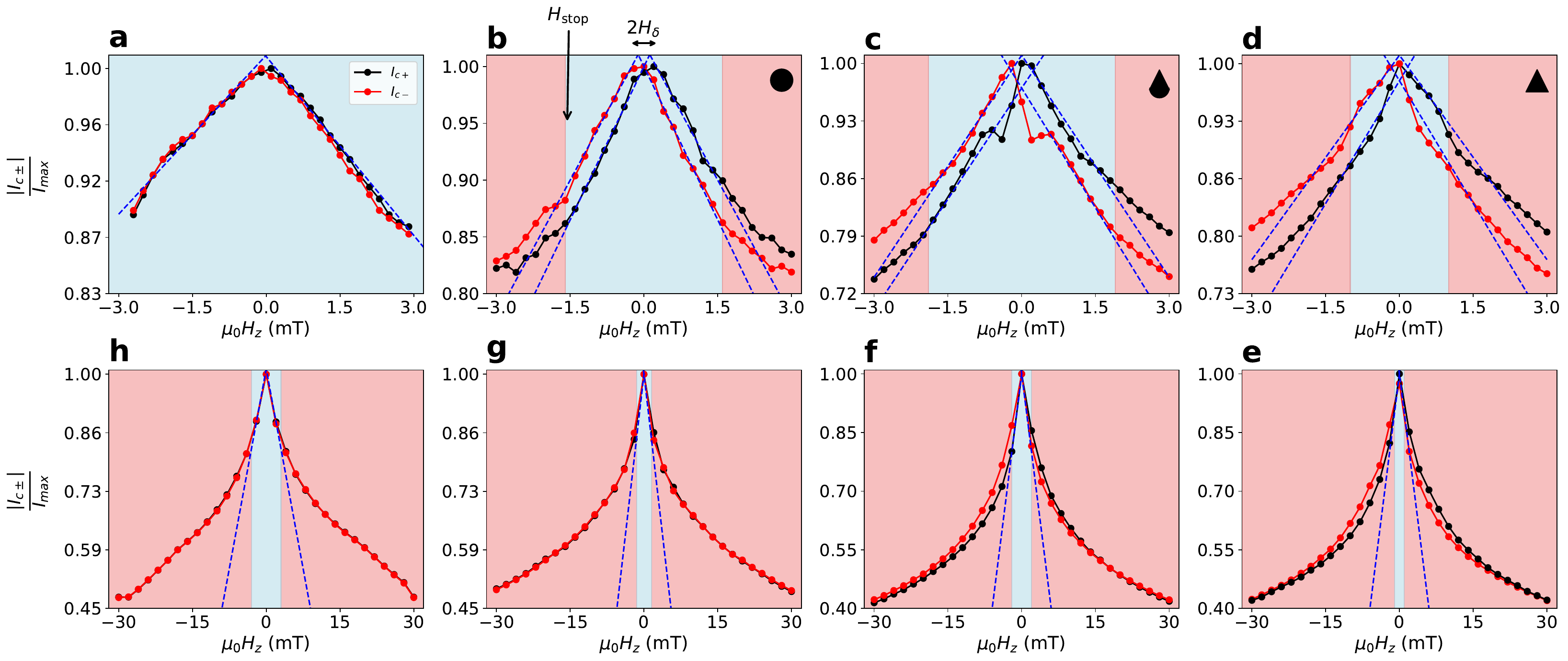}
    \caption{\textbf{Normalized critical currents across the four device geometries.} Normalized positive and negative critical currents ($I_c^\pm$) are shown for two magnetic field ranges: low ($\pm 3\,\mathrm{mT}$) and high ($\pm 30\,\mathrm{mT}$), for Reference (a,b), Hole (c,d), Drop (e,f), and Triangle (g,h) antidot devices. Corresponding maximum currents are $I_{\text{max}}^R = 1.3\,\mathrm{mA}$, $I_{\text{max}}^H = 1.9\,\mathrm{mA}$, $I_{\text{max}}^D = 2.8\,\mathrm{mA}$, and $I_{\text{max}}^T = 2.5\,\mathrm{mA}$. Measurements were performed at $T = 1.8\,\mathrm{K}$ for Hole, Drop, and Triangle, and $T = 1.9\,\mathrm{K}$ for Reference. 
    Blue shading indicates the Meissner state ($|H_z| < H_{\text{stop}}$), and red shading the mixed state. Linear fits in the Meissner regime (Eq.~\ref{eq.linear}) yield $\mu_0 H_{\text{stop}}^R = 2.6\,\mathrm{mT}$, $\mu_0 H_{\text{stop}}^H = 5.6\,\mathrm{mT}$, $\mu_0 H_{\text{stop}}^D = 5.0\,\mathrm{mT}$, and $\mu_0 H_{\text{stop}}^T = 3.5\,\mathrm{mT}$. Blue dashed lines represent linear fits from Eq.~\ref{eq.linear_asy}.}
    \label{all_samples}
\end{figure}

Figure~\ref{all_samples} compares the behavior of $I_c^\pm$ as a function of $H_z$ for the three antidot geometries and a reference device without antidots. We investigate two distinct field ranges, $\pm 3$~mT and $\pm 30$~mT, since the magnitude of $H_z$ determines the dominant depinning mechanism: surface depinning (driven by Meissner screening currents in low fields) or bulk depinning (prominent in higher fields). To facilitate comparison, all data are normalized to the maximum switching current."

Figures~\ref{all_samples}a,b show the evolution of the two critical currents for the reference device. Without an antidot, no SDE is observed throughout the field range within the resolution of the measurement ($I_c^+ \simeq I_c^-$), indicating negligible intrinsic asymmetry in the film.  
At low fields ($|\mu_0 H_z| \leq 5$~mT, Fig.~\ref{all_samples}a), the evolution of $I_c(H_z)$ is nearly linear, as expected from the Meissner screening that affects only the surface pinning and is in agreement with observations on similar device geometries~\cite{edgebarriers}. In
 higher fields ($|H_z| > 5$~mT), a sublinear dependence characterizes the damping of $I_c$. The transition between linear and sublinear regimes occurs at $\mu_0 H_{\text{stop}}^{\text{R}} \simeq 5$~mT, corresponding to $I_c(H_{\text{stop}}^{\text{R}})/I_c^{\text{max}} \simeq 0.75$, well above the 0.5 limit expected for a system with negligible bulk pinning~\cite{ic_p}. This confirms that bulk pinning plays a significant role in these Nb films even in the absence of antidots.

The evolution of $I_c(H_z)$ for antidot samples is very different. Indeed, as shown in fig.\ref{all_samples}c,d already with symmetric antidots (circles), a clear non-reciprocity of the supercurrent is visible ($I_c^+ \neq I_c^-$).
At low fields, $I_c(H_z)$ shows characteristic linear damping but with a clear and opposite shift in the field ($H_\delta\simeq \pm 0.2$mT) for $I_c^\pm$ as expected in the presence of edge asymmetries for flux depinning\cite{Moodera_ubq}. 
At larger fields ($\mu_0H_z \geq 2$mT, fig.\ref{all_samples}d), sublinear damping was observed with negligible SDE for the circular antidots, suggesting a recovered reciprocity in the presence of bulk pinning. 
By contrast, in the more asymmetric drop and triangle atidots, diode effects were visible in both field ranges.
At small fields (fig.\ref{all_samples}e,g), the linear damping is characterized by different slopes, indicating a richer dynamics of the supercurrent likely co-affected by additional crowding phenomena\cite{clem2,bridges, edgebarriers}. Indeed, while circular defects lack sufficiently pointed edges or relevant curvature angles to induce a current crowding effect, defects with a small radius of curvature or pointed geometries hinder the appearance of current crowding \cite{bridges,edgebarriers, clem1, georeduc}. 
Still, $H_\delta$ for all antidot devices is pretty similar, and also the average damping is compatible. At larger fields ($H_z>H_{\text{Stop}}$, red region in fig\ref{all_samples}f,h), the sublinear damping is also non-reciprocal, suggesting enhanced asymmetry also in the presence of bulk pinning. Notably, despite the complexity of the $I_c(H_z)$ behavior, the fundamental symmetry $I_c(H_z)^+=I_c(-H_z)^-$ is present for all devices, confirming the robustness of all the observed features.    

\begin{figure}[H]
    \centering
    \includegraphics[width=\linewidth]{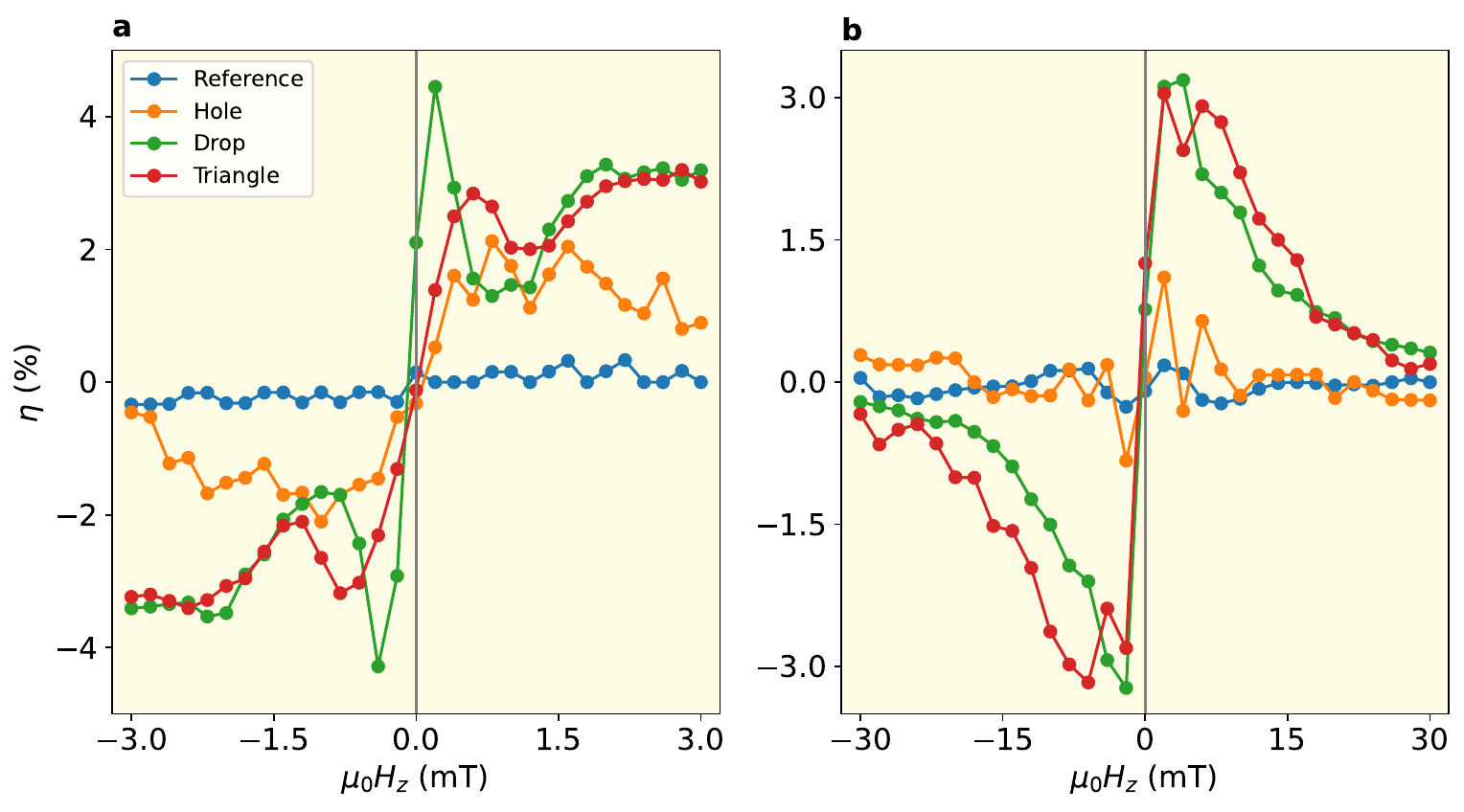}
    \caption{\textbf{SDE efficiency vs. out-of-plane magnetic fields.} Comparison of the field evolution of the diode efficiency $\eta$ for the four devices under an out-of-plane field $H_z$. The panels show field ranges of (a) 3~mT and (b) 30~mT.}
    \label{eta_bz}
\end{figure}

The SDE analysis can be further refined by comparing the diode efficiency, defined as $\eta \equiv \frac{I_c^+ - I_c^-}{I_c^+ + I_c^-},
$ for the four devices. In Fig.~\ref{eta_bz}, we report on the $\eta(H_z)$ evaluated over three different field ranges. At very small fields (Fig.~\ref{eta_bz}a), an almost linear increase of $\eta(H_z)$ is observed in all patterned devices, with more pronounced growth in the highly asymmetric dots and only a slight increase in the holes. Above $0.2~\mathrm{mT}$, the efficiency tends to saturate at a maximum value of about $4\%$ (see Fig.~\ref{eta_bz}b), and then decreases for $H_z > 10~\mathrm{mT}$, reaching negligible values above $30~\mathrm{mT}$. 
At large magnetic fields, only the triangular and drop-shaped samples exhibit a diode effect, whereas the symmetric dots do not, confirming the important interplay between shape and bulk pinning asymmetry at high fields.

\subsection*{Analytical Model of Vortex Dynamics}\label{subsec2}

\begin{figure}[hbt!]
    \centering
    \includegraphics[width=\linewidth]{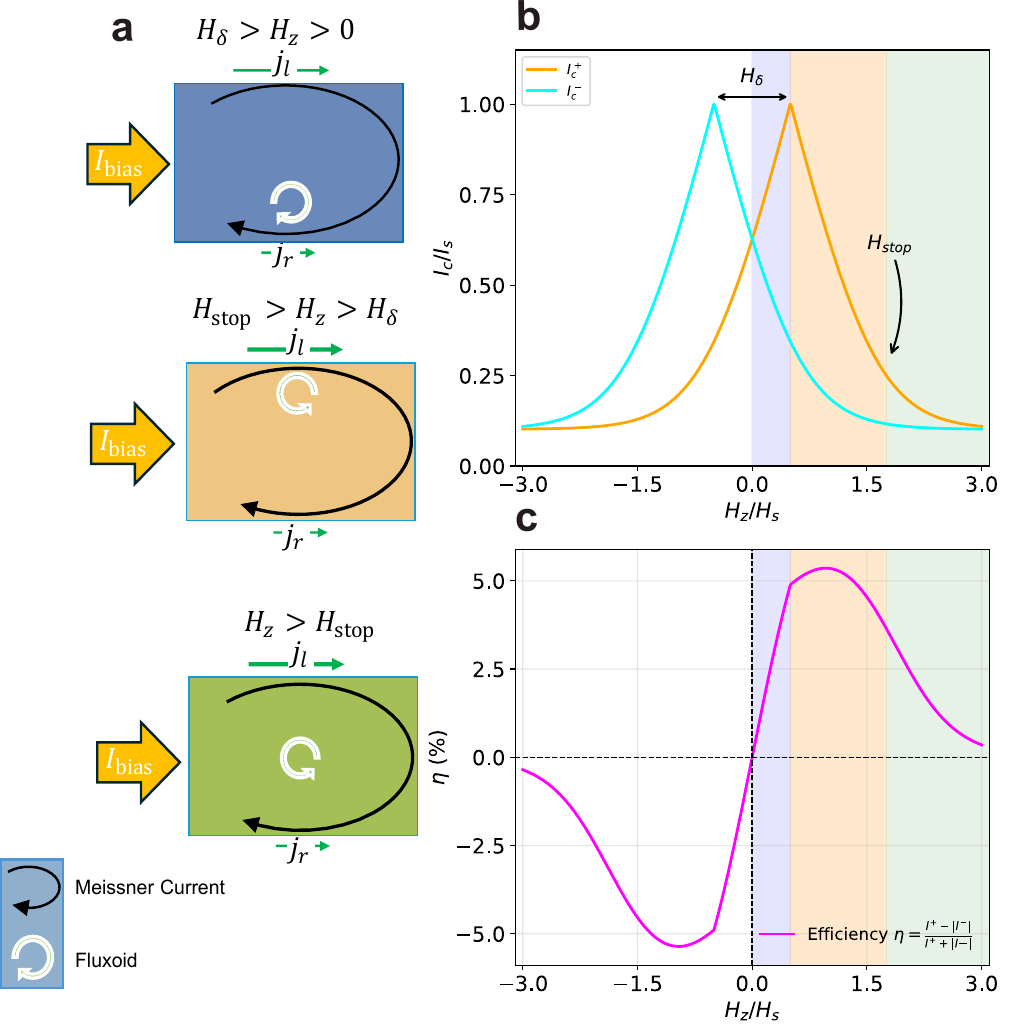}
    \caption{\textbf{Model for the SDE in the presence of an out-of-plane magnetic field.} 
    (a) Sketch of vortex dynamics in the presence of a bias current and Meissner screening currents in a film with different edge current thresholds ($j_s^r < j_s^l$). Three different field regimes are considered, corresponding to the intervals defined by $H_\delta$ and $H_{\text{stop}}$. 
    (b) $I_c^\pm$ evaluated from Eq.~\ref{eq.linear_asy} for $|H_z| < H_{\text{stop}}$ and using the system of equations from Ref.~\cite{ic_p} above $H_{\text{stop}}$. Model parameters: $p=0.2$ and $H_\delta=0.05 H_s$. 
    (c) Example of $\eta(H_z)$ computed from Eqs.~\ref{eq.eta_low}--\ref{eq.eta_high}. Parameters: $\eta_j=0.04$, $p=0.5$, $H_s=10$~mT, $\eta_p=-0.005$. In panels (b) and (c), the background colors represent the three relevant field ranges as described in (a).}
    \label{fig:theory}
\end{figure}
In superconducting thin films subjected to an out-of-plane magnetic field, the maximal dissipationless current that can flow in the films depends on a non-trivial interplay between the geometrical edge barrier and bulk vortex pinning. Indeed, above the critical current, the non-uniform supercurrent density flowing in the films induces the generation of vortices and antivortices at the two film edges. These can propagate and annihilate within the film, leading to energy dissipation and limiting supercurrent flow. For such vortex dynamics to occur, the applied current ($I$) must overcome two key thresholds: the edge or surface critical current ($I_{s}$) associated with the generation and depinning of the vortex at the edges of the film and the critical current of bulk pin ($I_p$) that governs the movement of the vortex across the film. The overall critical current of the film is thus bounded by the larger of these two values: $I_c \geq \max(I_p, I_{s})$ \cite{ic_p}.

The edge critical current can be expressed as $I_{s} \simeq  j_s \pi d \sqrt{2W\Lambda}$, where $d$ is the thickness of the film, $W$ is the width, $\Lambda$ is the Pearl length, and $j_s$ is the edge critical current density~\cite{ic_p}. For ideal edges, $j_s$ equals the Ginzburg-Landau depairing current density $j_{GL}$ \cite{aslamazov1983resistive}, while for realistic defected edges, it is typically smaller ($j_s < j_{GL}$)~\cite{ilin_critical_2014}.
It has been shown that at low magnetic fields ($ H_z<H_{\text{Stop}}$) and for $I_{s}<\frac{\pi}{2} I_p$, the dependency $I_c(H_z)$ can be linearized in the simple form \cite{ic_p}:
\begin{equation}
    I_c(H_z) = I_{s} - \left| H_z \right| 2\pi W,
    \label{eq.linear}
\end{equation}
In particular, this equation highlights that bulk pinning does not contribute at low fields and that the critical current monotonically decreases with increasing $H_z$. 
This reduction is due to the enhanced current density at one edge of the device arising from Meissner screening currents induced by the applied magnetic field and leading to a directional imbalance in the edge current distribution.
Bulk pinning impact in the range of validity of the linear approximation limiting it up to  $H_{\text{stop}}$ defined as\cite{ic_p}:
\begin{equation}
    H_{\text{stop}}\equiv  \frac{H_{s}}{2}(1-(\frac{\pi }{2}p)^2)
\end{equation}

where $p=I_p/I_{s}$ is the ratio between the bulk pinning current and the edge current, while $H_s\equiv I_s/2\pi W$ is the field at which vortices enter the strip with no applied current, then equal to the Bean-Livingston barrier field or to the lower critical field $H_{c1}$ in the absence of the former \cite{benkraouda_critical_1998}. 
Equation\ref{eq.linear} assumes equal barrier strengths at both edges of the film, i.e., $j_s^l = j_s^r$, where $j_s^l$ and $j_s^r$ are the threshold current densities at the left and right edges, respectively, relative to the direction of the applied current. However, if this symmetry is broken due to edge imperfections or intentional asymmetries ($j_s^l \neq j_s^r$), the positive and negative critical currents may differ ($I_c^+ \neq I_c^-$)\cite{Ando_Observation_2020}. In such cases, the linear dependence described earlier is modified as follows\cite{Vodolazov2005}:

\begin{equation}
    |I_c^\pm(H_z)| = I_{s}^{\max} - \left| H_z \mp H_\delta \right|2\pi W ,
    \label{eq.linear_asy}
\end{equation}
where, $I_{s}^{\max} = \max (j_s^l, j_s^r) \pi d \sqrt{2W\Lambda} = (H_{s}+|H_\delta|)2\pi W$, $H_\delta \equiv \frac{j_s^l - j_s^r}{2} \sqrt{\frac{\Lambda}{2W}}$ and $H_{s}$ has been extended to its mean value $H_{s} \equiv \frac{j_s^l + j_s^b}{2} \sqrt{\frac{\Lambda}{2W}}$.
The evolution of $I_c^+(H_z)$ is shown in Fig.~\ref{fig:theory}b together with a sketch of the vortex dynamics in  Fig.~\ref{fig:theory}a. 
From the figure is possible to see that at very low field ($0<H_z < H_\delta$) the Meisner screening current, opposed to the bias current at the weak edge, tends to enhance $I_c^+$ up to  $I_{s}^{\max}$ and the vortex is generated in the right edge assumed to be the weaker ($j_s^r<j_s^l$). Above such a field, the screening current is so large that the vortex generation occurs on the left edge, resulting in a linear damping of $I_c^+$ at intermediate fields ($H_{\delta}<H_z < H_{\text{stop}}$).  
Above $H_{\text{stop}}$, the vortex dynamics is governed by a nontrivial interplay between the bulk and the edge pinning, reducing the linear damping to a sublinear dependence on $H_z$, which cannot be captured by a single analytical expression but requires a solution of equations~\cite{ic_p}. The evolution of $I_c^-$ follows the same description, but in the dual time reversal $I_c^-(H_z)=I_c^+(-H_z)$.

From Eq.~\ref{eq.linear_asy} it is possible to express the diode efficiency $\eta (H_z) $ in therms of the edge asymmetry $\eta_j\equiv \frac{j^l-j^r}{j^l+j^r}=\frac{H_\delta}{H_{s}}$ for the two filed ranges:
\begin{equation}
    \eta_{\text{low}}(H_z) = \eta_j \frac{Hz}{H_\delta} \quad \text{for} \quad |H_z| < |H_\delta|
    \label{eq.eta_low}
\end{equation}

\begin{equation}
        \eta_{\text{med}}(H_z) = \eta_j \frac{\operatorname{sgn}(H_z)}{1 + |\eta_j| - |H_z/H_s|} \quad \text{for} \quad H_{\text{stop}}>|H_z| > |H_\delta|
        \label{eq.eta_med}
\end{equation}

In particular, for $|H_z| < |H_\delta|$, the diode efficiency is linear in $H_z$, and is bounded to the intrinsic edge asymmetry $\eta_j$ which is reached in $|H_z| = |H_\delta|$. 
For $|H_z| > |H_\delta|$ the efficiency continues to increase above $\eta_j$ and is bounded by the maximum value $2\eta_j$.

For fields exceeding $H_{\text{stop}}$, the limitation of the model does not allow an analytical formula of $\eta(H_z)$. However, in the regime of relatively large magnetic fields, where $H_z \gg p H_s$, the behavior of the critical current can be approximated as~\cite{ic_p}:
\begin{equation}
    I_c(H_z)\approx I_s (p + \frac{H_{s}}{4|H_z|})
    \label{eq_sublinear}
\end{equation}

Assuming a small asymmetry in both the bulk pinning current $\Delta I_p=I_p^+-I_p^-<<I_s$ and the edge current $\Delta I_s=I_s^+-I_s^-<<I_s$, an analytical expression for the diode efficiency can be derived in large fields:
\begin{equation}
    \eta_{\text{high}}(H_z)\approx 2\eta_j \frac{(2\Delta I_p / \Delta I_s)|H_z| +H_s }{4p|H_z|+H_s} \operatorname{sgn}(H_z)
    \quad \text{for} \quad |H_z| >> H_{\text{stop}}
    \label{eq.eta_high}
\end{equation}
In particular, at large magnetic fields, the efficiency $\eta(H_z)$ decays rapidly, within only a few multiples of $H_s$, while remaining bounded by $2\eta_j$.
In Fig.~\ref{fig:theory}c, we show $\eta(H_z)$ simulated using the experimental parameters. Despite the qualitative nature of our model, the simulation reproduces the main experimental trends: a sharp linear increase in $\eta$ at low fields, sublinear growth in the intermediate regime, and pronounced damping at high fields. 

\subsection*{Numerical Model of Vortex Dynamics}

To support the simplified analytical model presented earlier, we performed numerical simulations by solving the time-dependent Ginzburg-Landau equations (TDGL) using the open-source pyTDGL package~\cite{pytdgl}. These simulations allow us to analyze the vortex dynamics in the presence of an out-of-plane magnetic field and verify the device properties. Further details on the simulation parameters and approach are provided in the Methods (Section~\ref{sec6}).

\begin{figure}[H]
    \centering
    \includegraphics[width=\linewidth]{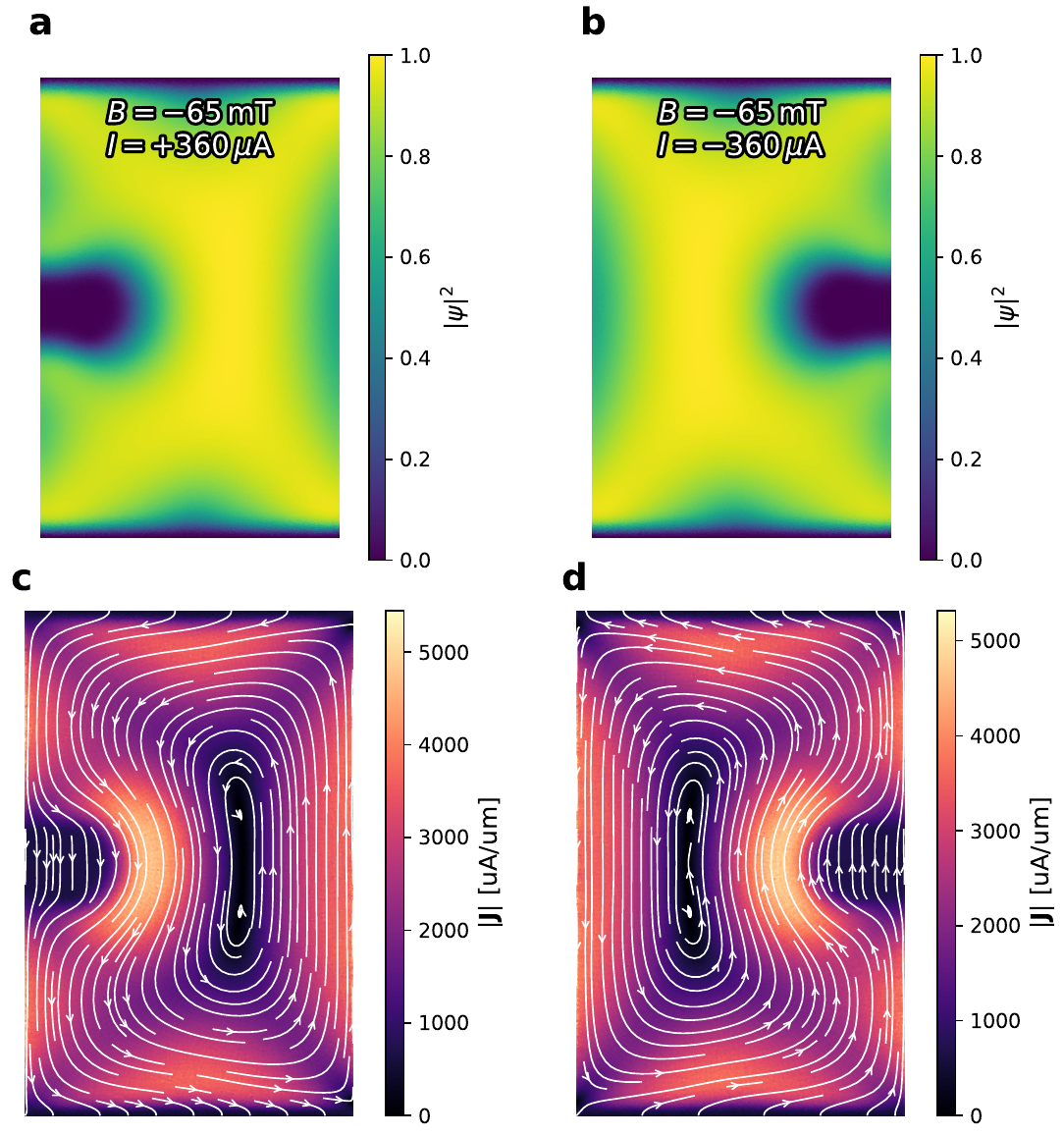}
    \caption{\textbf{TDGL simulation of the diode effect for the reference sample.} 
    (a, b) Order parameter distribution under an applied field of $B = -65$~mT after a simulation time of $50\tau$ for (a) positive and (b) negative bias currents. 
    (c, d) Corresponding current density distributions for the same conditions. Parameters used in TDGL simulations: coherence length $\xi = 0.9~\mu$m, penetration depth $\lambda = 1.35~\mu$m, and thickness $d = 0.2~\mu$m. These values satisfy the type-II superconductor condition $\kappa > 1/\sqrt{2}$.}
    \label{ref_sim}
\end{figure}

Figure~\ref{ref_sim} compares the spatial distribution of the order parameter (panels a and b) and the supercurrent densities (panels c and d), evaluated for positive and negative current biases at a fixed $H_z$. Figures~\ref{ref_sim}(a) and (b) show vortex nucleation occurring at opposite edges, consistent with theoretical expectations for Meissner screening. Specifically, for a positive current, vortices nucleate at the left edge, whereas for a negative current, they nucleate at the right edge. This confirms that current polarity determines which edge is susceptible to vortex entry. The current distribution maps (panels c and d) illustrate the behavior at both the center and edges of the superconductor; notably, when the energy barrier is suppressed and vortices enter, the edge current undergoes significant modification. Because the nucleation behavior is symmetric for both current polarities, this device does not exhibit an SDE, confirming the predictions of the analytical model. 

\begin{figure}[H]
    \centering
    \includegraphics[width=\linewidth]{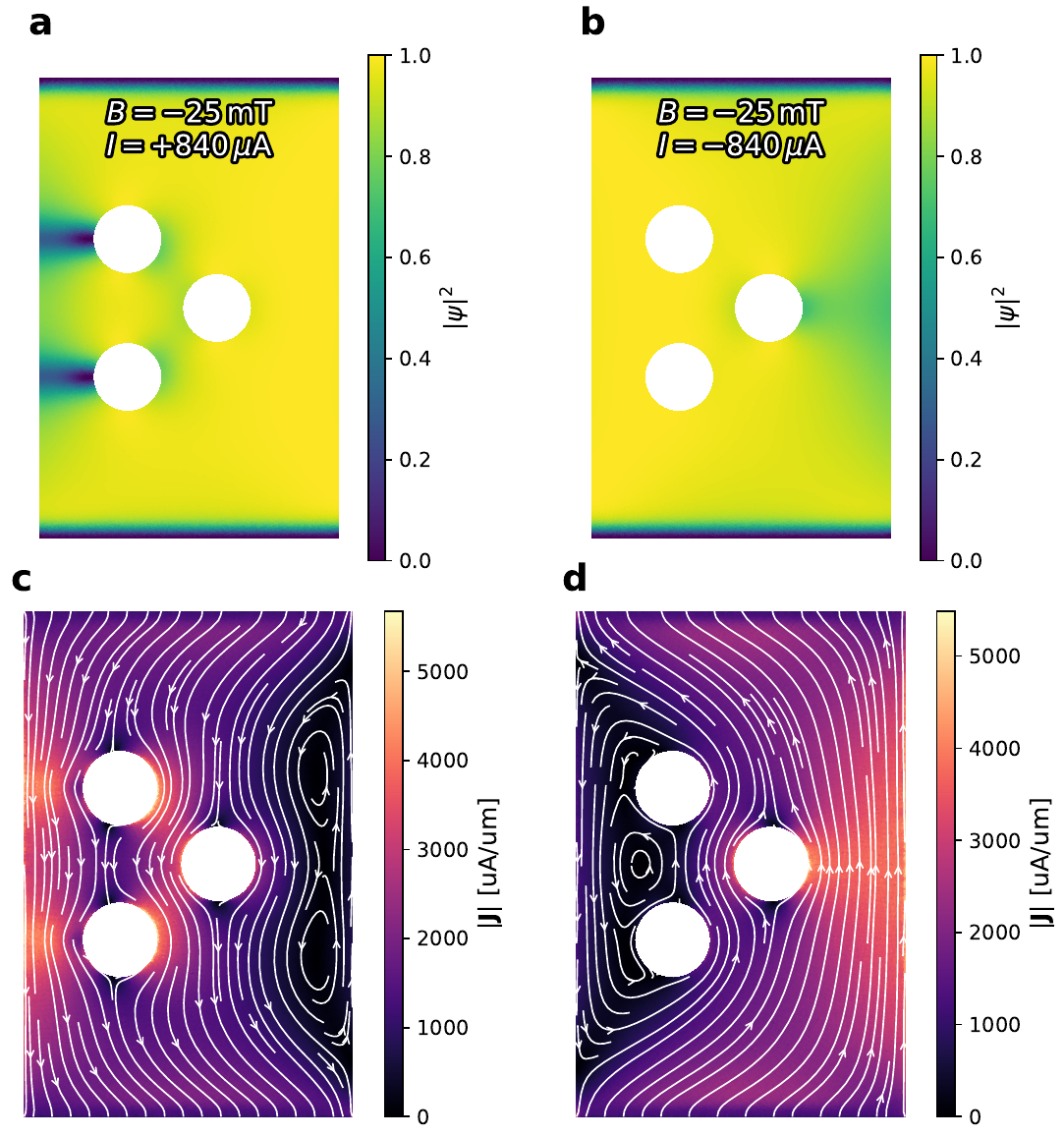}
    \caption{TDGL simulation of the diode effect from the Hole sample. a,b are the parameter order when B = -25mT is applied and after $50\tau$ of simulation to positive and negative current, respectively. c,d are the current distributions under the same conditions. Parameters used in TDGL simulations: $\xi = 0.9 \mu$m, $\lambda = 1.35 \mu$m and d = 0.2 $\mu$m. These parameter values guarantee the condition necessary for a superconductor type II $\kappa > 1/\sqrt{2}$.}
    \label{hol_sim}
\end{figure}

By contrast, the sample with circular antidots exhibits markedly different vortex dynamics, as shown in Fig.~\ref{hol_sim}. Due to the asymmetric positioning of the holes, which are closer to the left edge, vortex nucleation occurs in the region between the hole and the device boundary, consistent with theoretical predictions~\cite{edgebarriers, yu2007asymmetric}. Driven by current crowding, vortices nucleate at the hole edge and then propagate toward the device edge. 
The current distribution shows a strong concentration at this constriction, where the superposition of Meissner and applied currents increases the local density, thereby promoting nucleation. 
This behavior is particularly evident for the negative current polarity. Crucially, this phenomenology aligns with the analytical model, which incorporates an asymmetry in the edge-pinning threshold current densities at the left and right edges.

Similarly, for drop- and triangle-shaped devices, vortex nucleation consistently occurs at the defect tips, which are positioned to the left edge. Vortices nucleate at these tips and subsequently move toward the device boundary. 
This behavior is expected, as the pointed regions exhibit enhanced current density due to current crowding~\cite{clem1,bridges,cornershaped} arising from the superposition of Meissner and applied currents. Consequently, in the positive current configuration, the current density accumulates at the left edge, reducing the maximum positive critical current. In contrast, the backward supercurrent is enhanced by compensating for the Meissner screening current, which limits the edge current densities.

\begin{figure}[H]
    \centering
    \includegraphics[width=\linewidth]{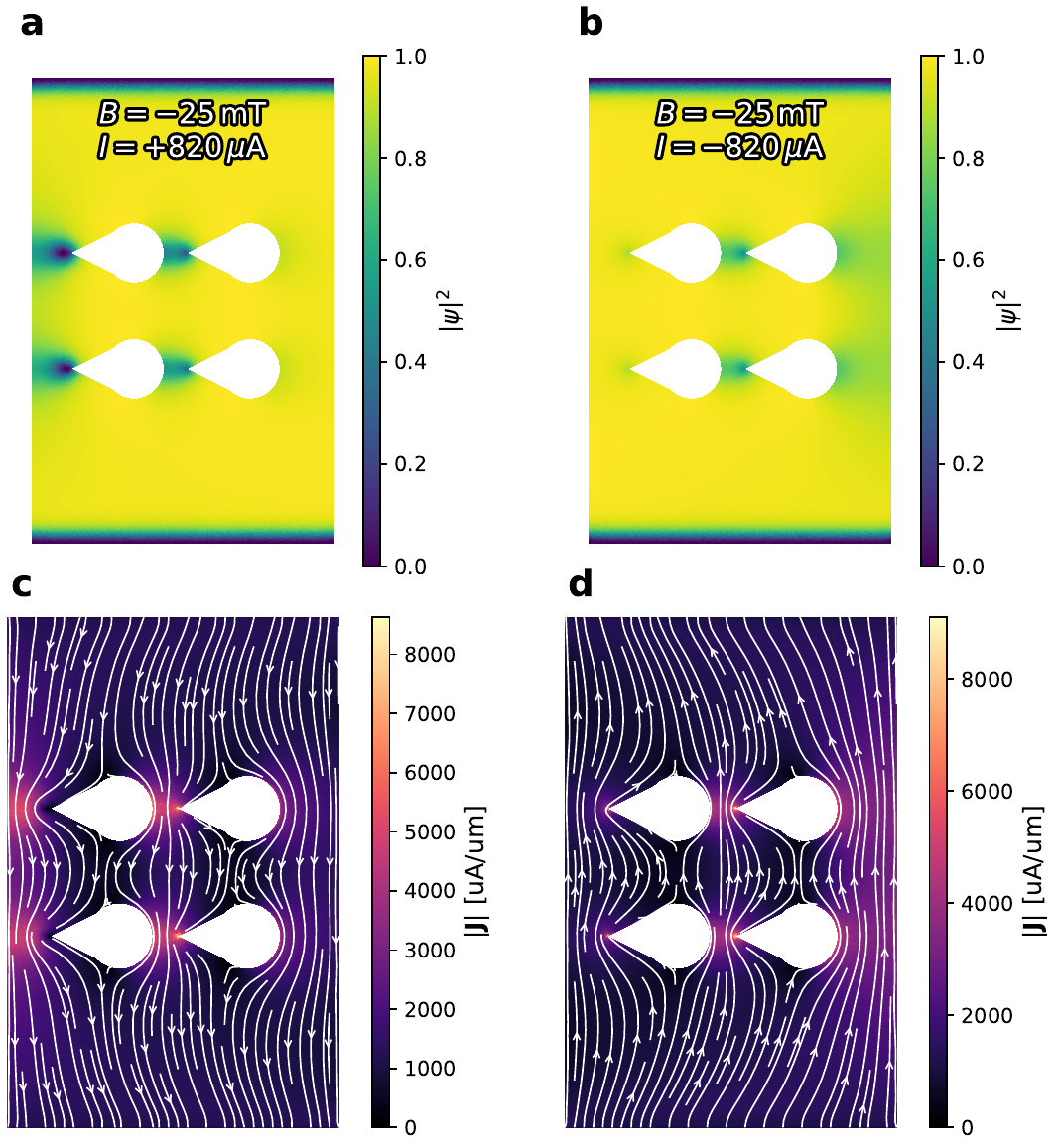}
    \caption{TDGL simulation of the diode effect from the Drop sample. a,b are the parameter order when B = -25mT is applied and after $50\tau$ of simulation to positive and negative current, respectively. c,d are the current distribution under the same conditions. Parameters used in TDGL simulations: $\xi = 0.9 \mu$m, $\lambda = 1.35 \mu$m and d = 0.2 $\mu$m. 
    These parameter values guarantee the condition necessary for a superconductor type II $\kappa > 1/\sqrt{2}$.}
    \label{drop_sim}
\end{figure}

Notably, because of computational constraints, the numerical simulations were performed on films with reduced dimensions and fewer antidots than in the experimental devices. 
This geometric rescaling justifies the use of larger magnetic-field ranges in the simulations than in the experiments. 
However, since our primary interest lies in the fundamental physical behavior and vortex dynamics, these differences in absolute values are not critical to qualitative conclusions.

\begin{figure}[H]
    \centering
    \includegraphics[width=\linewidth]{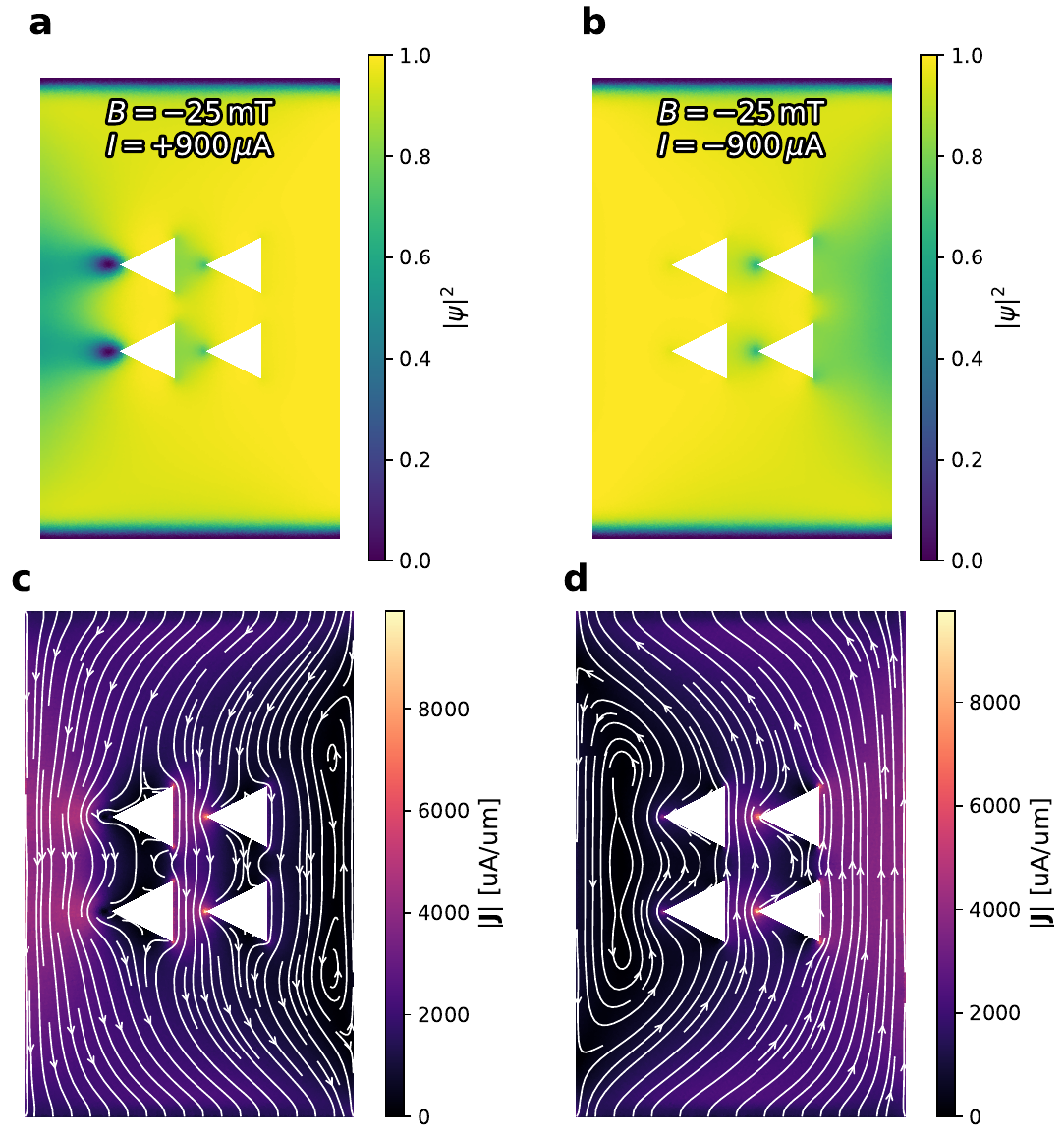}
    \caption{TDGL simulation of the diode effect from the Drop sample. a,b are the parameter order when B = -25mT is applied and after $50\tau$ of simulation to positive and negative current, respectively. c,d are the current distribution under the same conditions. Parameters used in TDGL simulations: $\xi = 0.9 \mu$m, $\lambda = 1.35 \mu$m and d = 0.2 $\mu$m. These parameter values guarantee the condition necessary for a superconductor type II $\kappa > 1/\sqrt{2}$.}
    \label{tri_sim}
\end{figure}


\subsection*{In-plane magnetic field response.}\label{subsec2}

The behavior of the critical current ($I_{\mathrm{c}}$) and the resulting SDE under an in-plane magnetic field ($H_{\mathrm{in}}$) differs markedly from the out-of-plane case. As expected, contrary to the strong suppression observed with a perpendicular field $H_z$, the suppression of $I_{\mathrm{c}}$ by $H_{\mathrm{in}}$ is much weaker. This is attributed to reduced Meissner screening currents and only partial penetration of the in-plane magnetic field. Nevertheless, at relatively large magnetic fields ($H_{\mathrm{in}} > 10~\mathrm{mT}$), a finite SDE is observed in all devices, including the reference sample. In particular, since the film thickness $d$ exceeds the critical threshold $d > 1.84 \xi_s$~\cite{fink_vortex_1969} and the measurement temperature is above $T_c/2$~\cite{vlasko-vlasov_crossing_2016}, the nucleation of ordered vortices in the plane is expected. Consequently, the dynamics governing the SDE may share significant similarities with the out-of-plane regime.

\begin{figure}[hbt!]
    \centering
    \includegraphics[width=\linewidth]{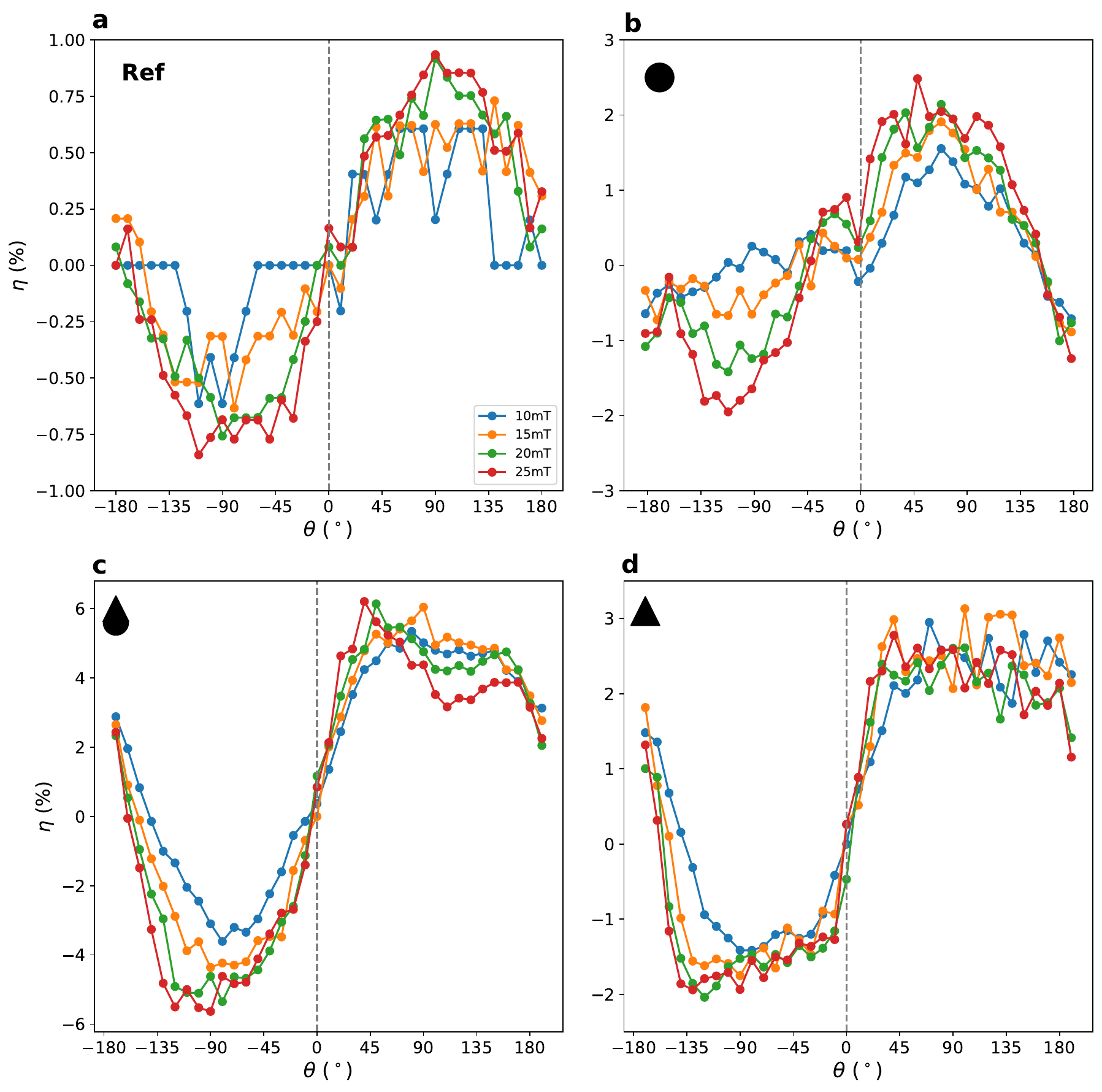}
    \caption{\textbf{In-plane angular dependence of SDE efficiency.} Full scan of the SDE efficiency as a function of $\theta$ (the orientation of the in-plane magnetic field $B_{\mathrm{in}}$ relative to the supercurrent direction). Panels show the reference device (a), the hole (b), the drop (c), and the triangle (d) antidot devices. The magnitude of $B_{\mathrm{in}}$ ranges from 10 to 25\,mT.
        }
    \label{eta_Binplane}
\end{figure}

Figure~\ref{eta_Binplane} compares the diode efficiency measured for various in-plane magnetic field amplitudes (up to 25~mT) and orientations relative to the bias current. The field orientation is defined by the angle $\theta$, where $\theta = 0^{\circ}$ corresponds to a field parallel to the current and $\theta = 90^{\circ}$ to an orthogonal configuration. Across all devices, the diode efficiency is maximized when the magnetic field is perpendicular to the current ($\theta = \pm 90^{\circ}$) and becomes negligible in the parallel configuration. This angular dependence is consistent with the conventional SDE mechanism driven by magnetochiral anisotropy.

\begin{figure}[hbt!]
    \centering
    \includegraphics[width=\linewidth]{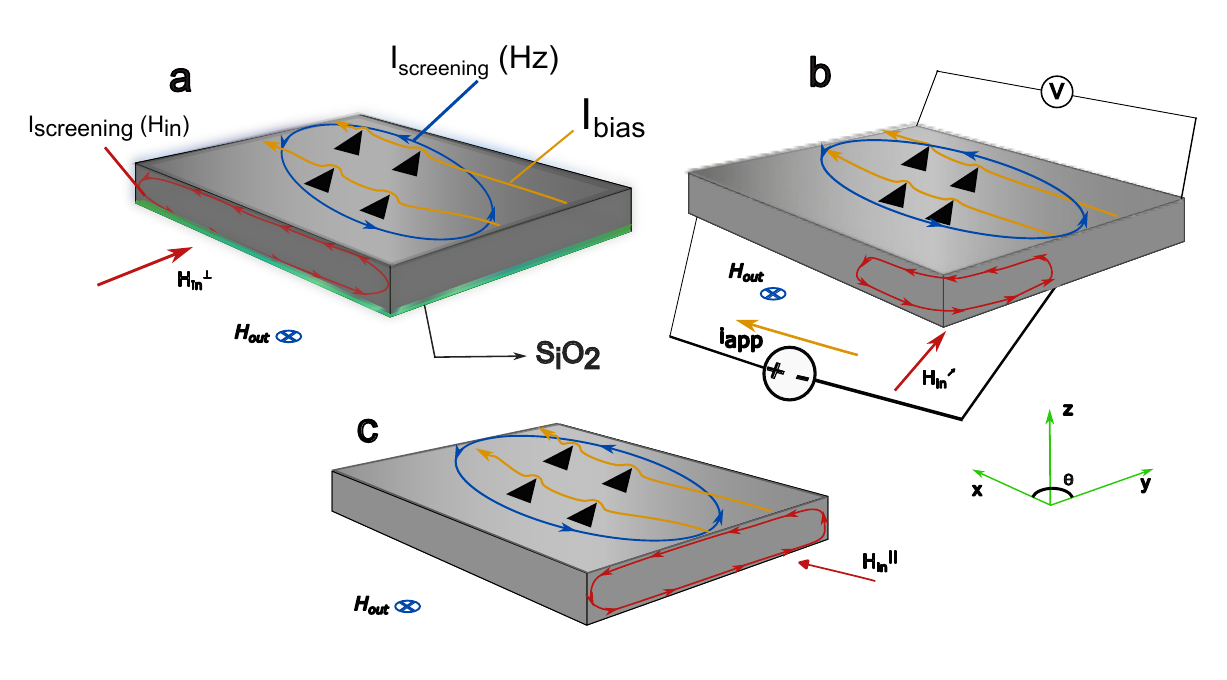}
    \caption{\textbf{Current distribution schematics for varying in-plane field orientations.} 
Superposition of the applied transport current (orange), $H_{z}$-induced Meissner currents (blue), and $H_{\mathrm{in}}$-induced currents (red). The transport current flows along the $x$-axis. The in-plane field $B_{\mathrm{in}}$ is applied: 
\textbf{(a)} along the $y$-axis (perpendicular to current); 
\textbf{(b)} along a diagonal direction; and 
\textbf{(c)} along the $x$-axis (parallel to current). 
Panel (a) highlights the Nb/SiO$_2$ interface, which is present in all devices. 
}
    \label{esq_Bin}
\end{figure}

In the reference device, the overall efficiency is lower, yet the angular dependence remains smooth and sinusoidal. 
By contrast, antidot devices exhibit a more irregular angular dependence. This behavior is consistent with a diode effect driven by Meissner screening currents and asymmetric surface barriers at the top (native oxide) and bottom (substrate) interfaces. This phenomenology mirrors the SDE observed in the previous section for low out-of-plane magnetic fields. The deviations of $\eta(\theta)$ from an ideal sinusoidal profile in antidot devices can be attributed to the geometrical effects introduced by the antidots. Specifically, the antidots may distort the magnetic field lines and induce current crowding within the screening current streamlines, thereby enhancing the diode effect even for non-orthogonal field orientations (see the sketch in Fig.~\ref{esq_Bin}).

At larger in-plane magnetic fields (\(B_{\mathrm{in}} > 25~\mathrm{mT}\)), the diode effect remains observable; however, special care must be taken to exclude spurious out-of-plane magnetic-field components arising from a misalignment of the vector magnet with respect to the device plane. For example, a misalignment as small as \(1^{\circ}\) can result in an effective perpendicular field \(H_z \simeq 2~\mathrm{mT}\) at \(B_{\mathrm{in}} = 100~\mathrm{mT}\), which is sufficient to induce a sizable out-of-plane SDE.

\begin{figure}[hbt!]
    \centering
    \includegraphics[width=\linewidth]{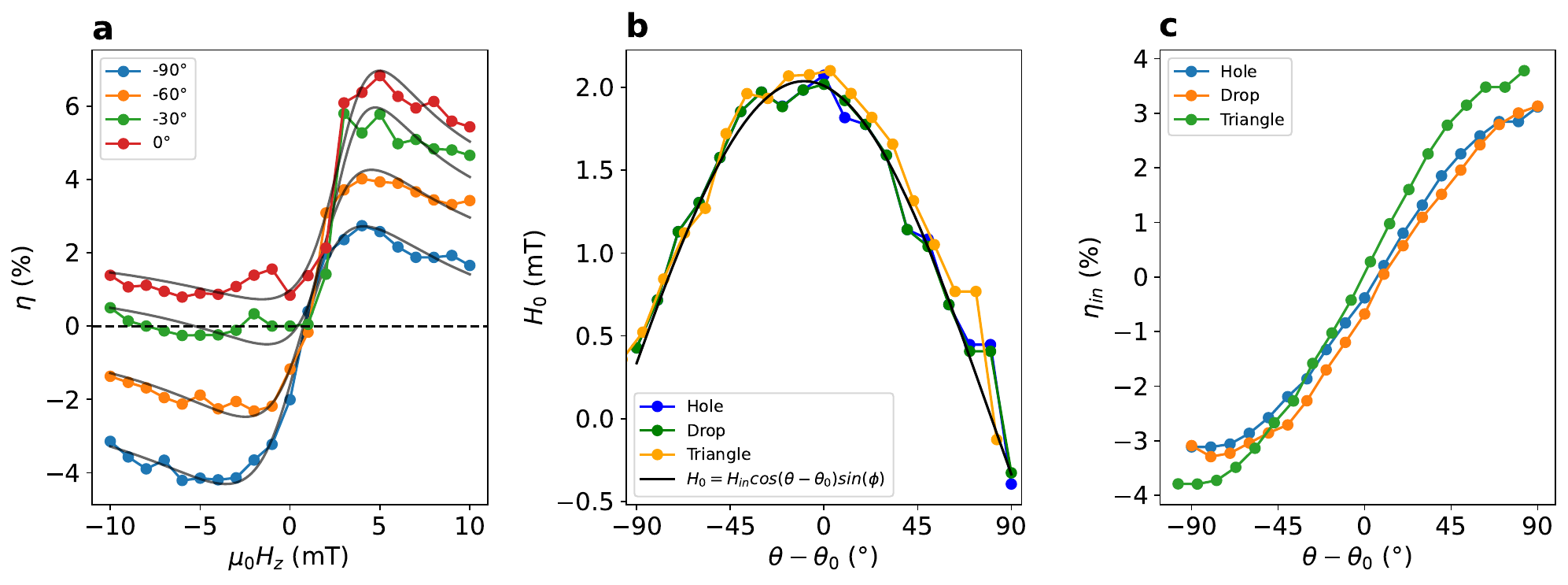}
    \caption{\textbf{Angular evolution of the SDE at large in-plane fields.} 
(a) Efficiency of the hole antidot device as a function of $H_z$ for different in-plane angles $\theta$, with $\mu_0 H_{\mathrm{in}} = 100$\,mT. The data are fitted using the model in Eq.~\ref{Eq.H_0} (solid line) with the following parameters: $a$ ranging from 1--3\,mT$^{-1}$ and $b$ ranging from 0.1--0.5\,mT$^{-2}$. 
(b) The parameter $H_0(\theta-\theta_0)$ resulting from the fitting procedure for the three antidot devices. Solid lines indicate fit to Eq.~\ref{Eq.H_0} with parameters $\phi = 1.2^{\circ}$ and $\theta_0^H = \theta_0^D= -9.5^{\circ} $, $\theta_0^T = -9.5^{\circ} $. 
(c) SDE efficiency $\eta_{\mathrm{in}}$ as a function of $\theta$ extracted from the fits.
}
    \label{etain_high}
\end{figure}

To quantitatively assess and eliminate this contribution, for each angle \(\theta\) a small scan of \(H_z\) was performed in a fixed in-plane field. 
Figure~\ref{etain_high}a shows a representative example of the hole antidot device, where the efficiency of the diode \(\eta(H_z)\) is measured for several values of \(\theta\). In addition to a constant offset \(\eta_{\mathrm{in}}\), attributed to the in-plane magnetic field, an asymmetric modulation of \(\eta(H_z)\) is observed. This modulation closely resembles the behavior of \(\eta(H_z)\) measured in the absence of an in-plane field, as discussed in the previous section, but with the symmetry center shifted away from the zero field.
The total efficiency of the diode $\eta(H_z)$ is then consistent with an additive contribution from the two applied fields and can therefore be written as
\begin{equation}
    \eta(H_z,\theta) = \eta_{\mathrm{in}}(\theta) + 
    \eta_{z}\bigl(H_z - H_0(\theta)\bigr),
\end{equation}
where the field offset \(H_0(\theta)\) accounts for the effective perpendicular component generated by a misalignment angle \(\phi\) between the sample plane and the \(x\text{--}y\) plane of the vector magnet:
\begin{equation}
    H_0(\theta) = H_{\mathrm{in}} 
    \cos(\theta - \theta_0)\sin\phi .
    \label{Eq.H_0}
\end{equation}
The out-of-plane contribution \(\eta_z\) can be fitted using the simplified analytical expression
\begin{equation}
    \eta_z(H_z - H_0) = 
    \frac{a (H_z - H_0)}{1 + b (H_z - H_0)^2},
    \label{eq.fit}
\end{equation}
where \(a\) and \(b\) are fitting parameters. This functional form captures the linear increase of \(\eta_z\) in low fields and the damping of \(1/H\) in larger fields, in agreement with the behavior discussed in the previous section and in good agreement with the experimental data shown in fig.~\ref{etain_high}a.

Figures~\ref{etain_high}b and~c show the extracted \(H_0(\theta)\) and \(\eta_{\mathrm{in}}(\theta)\) for the three antidot devices. The angular dependence of \(H_0(\theta)\) follows the expected trigonometric form for all devices depicted in eq.\ref{Eq.H_0}, resulting in a consistent misalignment angle of \(\phi \simeq 1.2^{\circ}\) and $theta_0$ dependent on the relative orientation of each device. Moreover, Fig.~\ref{etain_high}c demonstrates that at large in-plane magnetic fields the behavior of \(\eta_{\mathrm{in}}(\theta)\) becomes more regular across the antidot devices, evolving toward a quasi-ideal sinusoidal dependence. In this regime, a maximum diode efficiency of up to \(4\%\) is achieved, consistent with a reduced distortion of the magnetic field at higher \(H_{\mathrm{in}}\).

\subsection*{Impact of In-Plane Magnetic Field on SDE in the Vortex State}\label{subsec3}

The impact of an in-plane magnetic field on the SDE is now investigated in the presence of a high out-of-plane field ($H_z$). In this regime, the Nb film is populated by Abrikosov vortices, and the low-field SDE typically vanishes, as demonstrated in the previous section. This condition enables the study of SDE in the presence of vortices, while corrections arising from minor misalignments between the in-plane field and the device plane can be neglected.

\begin{figure}[hbt!]
    \centering
    \includegraphics[width=\linewidth]{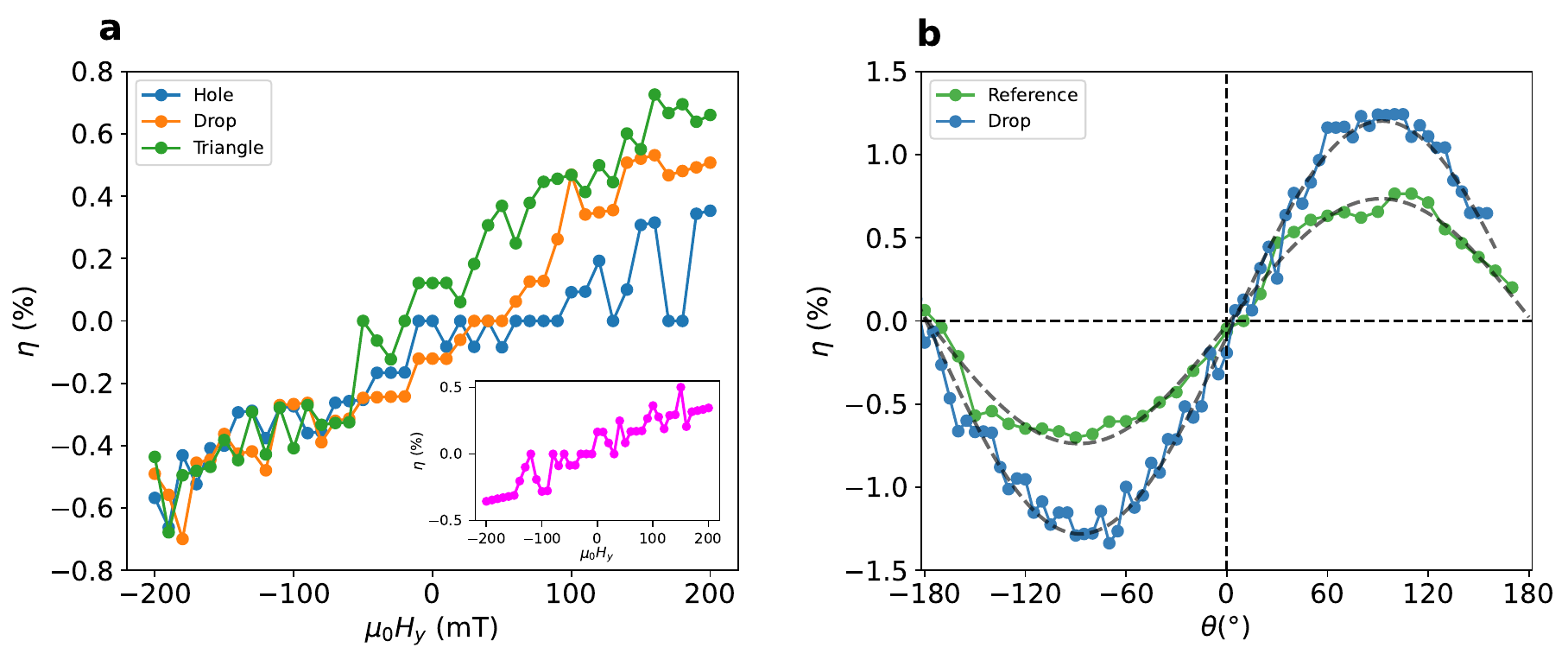}
    \caption{\textbf{In-plane SDE in the Abrikosov vortex state.}  
    \textbf{a,} Efficiency $\eta(H_y)$ measured for the three antidot devices at a constant out-of-plane field $H_z=50$~mT. The inset displays $\eta(H_y)$ for the drop device at $H_z=-50$~mT. 
    \textbf{b,} Angular dependence of the SDE efficiency $\eta$ measured for the reference and drop devices. The data were shifted angularly to correct for physical misalignment between the samples, with $\theta$ defined relative to the bias current direction. The Drop device shows a misalignment of $ \theta \approx 20^{\circ}$ during the measurement. The measurements were performed at temperatures of 1.8 K and 1.9 K for the Drop and Reference devices, respectively.}
    \label{HighFileds}
\end{figure}

$I_{c}^+$ and $I_{c}^-$ on high $H_z$ for the antidot devices decrease as $H_z$ increases, a behavior consistent with vortex entry and isotropic bulk pinning, while no diode effect was observable in this field range. However, upon applying an in-plane magnetic field orthogonal to the current direction ($H_y$), non-reciprocity re-emerges even in this high-field regime ($H_z=50$~mT). In this configuration, the SDE efficiency $\eta(H_y)$ scales linearly across the explored range $\pm 200$~mT (Fig.~\ref{HighFileds}b). This behavior aligns with the Meissner screening hypothesis (Eq.~\ref{eq.eta_low}) and confirms that the magnetic fields governing the SDE are additive. In contrast to the out-of-plane field behavior, where the linear region is limited to a few mT (see fig.~\ref{eta_bz}), $\eta(H_y)$ exhibits a much broader linear range. This observation is consistent with the screening model described in Eq.~\ref{eq.eta_low}. In the case of in-plane fields, the relevant geometric dimension that governs the screening shifts from the device width $W$ to the thickness of the film $d$. Since the thickness is much smaller than the width ($d \ll W$), the characteristic field scale $H_\delta$, which scales as $d^{-1/2}$, increases significantly compared to the out-of-plane configuration (where $H_\delta \propto W^{-1/2}$).

The polarity of $\eta(H_y)$ remains unchanged by reversing $H_z$ as shown in the inset of fig.~\ref{HighFileds}b, confirming that the observed SDE is independent of the Abrikosov vortex polarity.

Figure~\ref{HighFileds}c compares the angular dependence $\eta(\theta)$ of the antidot device with that of the reference sample. Both devices exhibit a sinusoidal trend, confirming that even in this regime, the SDE is governed solely by the magnetic field component orthogonal to the current direction. In particular, the antidot sample demonstrates a fourfold increase in efficiency compared to the reference. This increase is attributed to current crowding phenomena induced by constrictions around antidots, consistent with observations at low $H_z$. 
However, it should be noted that the overall SDE efficiency in the presence of Abrikosov vortices is reduced compared to the vortex-free regime ($H_z=0$) described in the previous section.

\section*{Conclusions}\label{sec13}

In conclusion, we have demonstrated that the SDE can be engineered in thin Nb films by combining asymmetric antidot geometries with magnetic fields. Our vectorial field analysis reveals that asymmetric flux pinning governs the SDE for both in-plane and out-of-plane magnetic field orientations.

We observed that the films are highly sensitive to out-of-plane magnetic fields ($H_z$). A comparison with a pristine reference device confirms that antidot devices, even symmetric circular holes, can induce SDE in films where it is otherwise absent, due to the broken inversion symmetry introduced by their spatial arrangement~\cite{Vodolazov2005, nanoholes}. By contrast, an in-plane magnetic field ($H_y$) orthogonal to the supercurrent induces SDE in all devices, including the pristine reference. This behavior is attributed to the surface barrier asymmetry between the top (NbO$_x$) and bottom (SiO$_x$) interfaces.

Notably, the SDE efficiencies from these two mechanisms are additive, indicating that in-plane and out-of-plane vortex nucleations are additive phenomena. 
In the high-field regime (presence of Abrikosov vortices), the in-plane SDE becomes less efficient but more predictable, exhibiting a linear dependence on $H_y$ and sinusoidal in the angle $\theta$ between the current and the field direction.

We successfully described these findings using an analytical model accounting for both edge and bulk pinning. In low fields, the linear increase of $\eta$ with $H_y$ and $H_z$ arises from the interplay between asymmetric edge pinning and Meissner screening. At high $H_z$, bulk pinning also becomes relevant, sustaining the SDE primarily in devices with strongly asymmetric antidots. 
The model further predicts that the maximum diode efficiency is limited by twice the edge asymmetry factor $2\eta_j$. Remarkably, this implies that near-ideal efficiencies ($\eta \approx 1$) could be achieved even with non-ideal edge asymmetries by operating near the stopping field $H_{\text{stop}}$. 

These results highlight the versatility of flux-based superconducting diodes controlled by vectorial fields and pave the way for future optimized designs~\cite{Mehrnejat2024, Nadeem2023, Ma2025}. In particular, hybridization with ferromagnetic films~\cite{Moodera_ubq} offers a promising route to implement these concepts in energy-harvesting or detection technologies, analogous to recent advances in superconducting tunnel diodes~\cite{strambini_superconducting_2022, geng_superconductor-ferromagnet_2023, araujo_superconducting_2024}. Furthermore, the control of diode polarization via magnetic fields paves the way for superconducting logic, neuromorphic computation~\cite{borgongino_biharmonic-drive_2025,hosur_digital_2024,gupta_gate-tunable_2023}, and memories~\cite{golod_demonstration_2022}. Although research in this area is still in its infancy, these developments mark a significant step toward the comprehensive understanding and application of SDEs.

\backmatter

\section*{Methods}\label{sec5}

\subsection*{Sample fabrication and measurement setup} \label{sec6}

For sample fabrication, the positive e-beam resist ARP 672.08 was spin-coated at 4000~RPM for 60~s and soft-baked at 150~$^{\circ}$C for 180~s. The resist was exposed using electron-beam lithography (Tescan Mira) and developed in a solution of IPA:DIW (7:3) for 120~s, followed by a rinse in DIW for 60~s and drying with N$_2$. 
Nb films (200~nm thick) were deposited using a DC Magnetron Sputtering System (AJA) with a base pressure of $2\times10^{-7}$~Torr and a working pressure of $6\times10^{-4}$~Torr. 
Pattern transfer was achieved by lift-off in acetone at 50~$^{\circ}$C for 300~s, followed by rinsing in IPA for 30~s and drying with N$_2$. 

The samples were wire-bonded with aluminum wires and magnetoelectric measurements were performed in a dilution cryostat equipped with a vector magnet. In all measurements, the signals were amplified using low-noise voltage preamplifiers. To improve measurement quality and ensure thermal stability, the RC filtering stages were minimized to reduce the power load that would otherwise cause excessive heating of the cryostat mixing chamber.

\subsection*{Numerical Modeling via Time-Dependent Ginzburg-Landau Theory}\label{sec7}

The investigation focused on the impact of geometric symmetry breaking on vortex dynamics and the superconducting diode effect (SDE). To this end, four distinct scenarios were modeled: a reference device with a perfectly smooth surface and three variants incorporating distinct surface defects (hole, drop, and triangle). These defects act as nucleation and pinning centers, modifying surface energy barriers and promoting non-homogeneous current distributions. 

Simulations were conducted using the open-source \texttt{pyTDGL} package, employing the finite volume method on an unstructured Delaunay triangular mesh~\cite{pytdgl}. To ensure numerical stability and accurate resolution of the vortex cores, the mesh density was configured with a maximum edge length of $\xi/2$. The applied vector potential was introduced as a gauge-invariant boundary condition, allowing for the study of vortex-antivortex pair nucleation and the system's critical current.

The dynamic properties and nonlinear magnetic response of the niobium (Nb) device were investigated using the generalized time-dependent Ginzburg-Landau (gTDGL) theory~\cite{pytdgl}. The model describes the temporal evolution of the complex order parameter $\psi$ and the electric scalar potential $\mu$ through numerically solved coupled partial differential equations in a two-dimensional domain. 
For material calibration, we adopted phenomenological parameters representing niobium in a temperature regime close to its critical temperature ($T \approx 0.998 T_c$)~\cite{squid_vortex}. At this temperature, the characteristic lengths diverge, thereby validating the thin-film (2D) approximation, in which the film thickness $d$ is much smaller than the coherence length. We assumed a coherence length $\xi = 0.9~\mu$m, a penetration depth $\lambda=1.35~\mu$m, and a thickness $d=0.2~\mu$m~\cite{squid_vortex}.

To optimize computational efficiency and facilitate scaling analysis, a global rescaling was applied to both geometric dimensions of the device and the characteristic lengths $\xi$ and $\lambda$ \cite{squid_vortex}. This procedure preserves the Ginzburg-Landau parameter ($\kappa=\lambda/\xi\approx1.5$) and the fundamental properties of a Type-II superconductor in the weak screening limit. 
In particular, the parameter $\gamma$ was kept invariant to maintain the consistency of the intrinsic relaxation time scale of the material, $\tau_0=\mu_0\sigma\lambda^{2}$ \cite{pytdgl,squid_vortex}.

\section*{ Data availability} 
Requests should be directed to Eloi Benicio.


\bibliography{sn-bibliography}

\section*{Acknowledgments}

The authors thank the Brazilian agencies CNPq 312697/2023-6, FAPEMIG APQ-04548-22, and Coordenação de Aperfeiçoamento de Pessoal de Nível Superior (CAPES)—Finance Code 001—for their financial support.
F.G. and E.S acknowledge the PNRR MUR project PE0000023-NQSTI for partial financial support.
E.S acknowledges the partial support from the Italian National Research Council (CNR) under grant DFM.AD002.206 (HELICS)

\section*{Competing interests}
The authors declare no competing interests.

\section*{ Author contribution}

E.B.M.J., C.I.L.A., and E.S. conceived, conducted, and analyzed the experiments. 
E.B.M.J. fabricated the samples. E.B.M.J., C.I.L.A., E.S., and F.G. wrote the paper.

\bigskip
\begin{flushleft}%
Editorial Policies for:

\bigskip\noindent
Springer journals and proceedings: \url{https://www.springer.com/gp/editorial-policies}

\bigskip\noindent
Nature Portfolio journals: \url{https://www.nature.com/nature-research/editorial-policies}

\bigskip\noindent
\textit{Scientific Reports}: \url{https://www.nature.com/srep/journal-policies/editorial-policies}

\bigskip\noindent
BMC journals: \url{https://www.biomedcentral.com/getpublished/editorial-policies}
\end{flushleft}

\section{Supplementary Information}\label{secA1}

\subsection{Film Critical Temperature and Field}

\begin{figure}[hbt!]
    \centering
    \includegraphics[width=\linewidth]{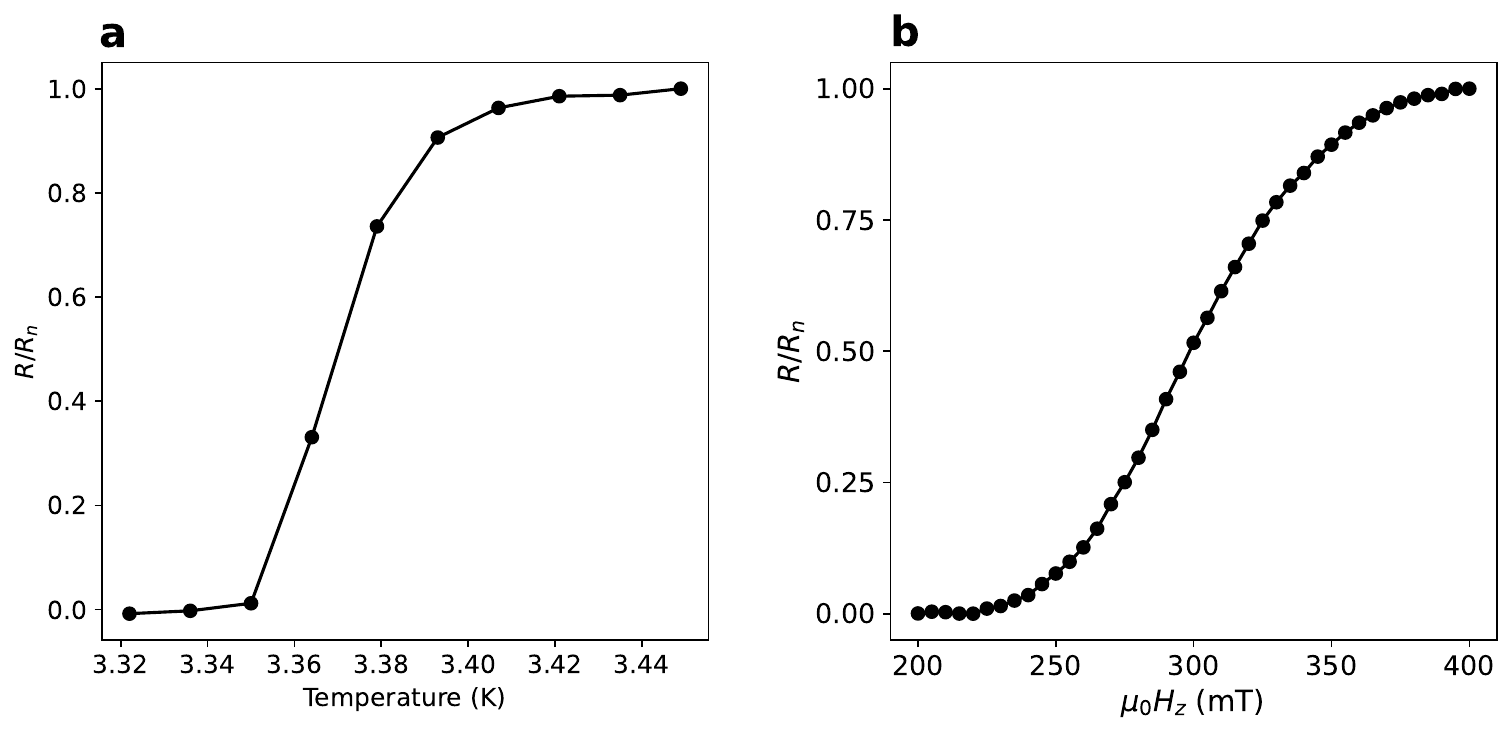}
    \caption{\textbf{Superconducting transitions of the Nb film.} \textbf{a}, Resistance as a function of temperature, yielding a critical temperature $T_c \simeq 3.37$~K. \textbf{b}, Resistance versus out-of-plane magnetic field, showing an upper critical field $H_{c2} \simeq 300$~mT. Both $T_c$ and $H_{c2}$ were extracted using the $50\%$ normal-state resistance criterion ($R = 0.5 R_n$).}
    \label{hc2}
\end{figure}

\begin{figure}[H]
    \centering
    \includegraphics[width=\linewidth]{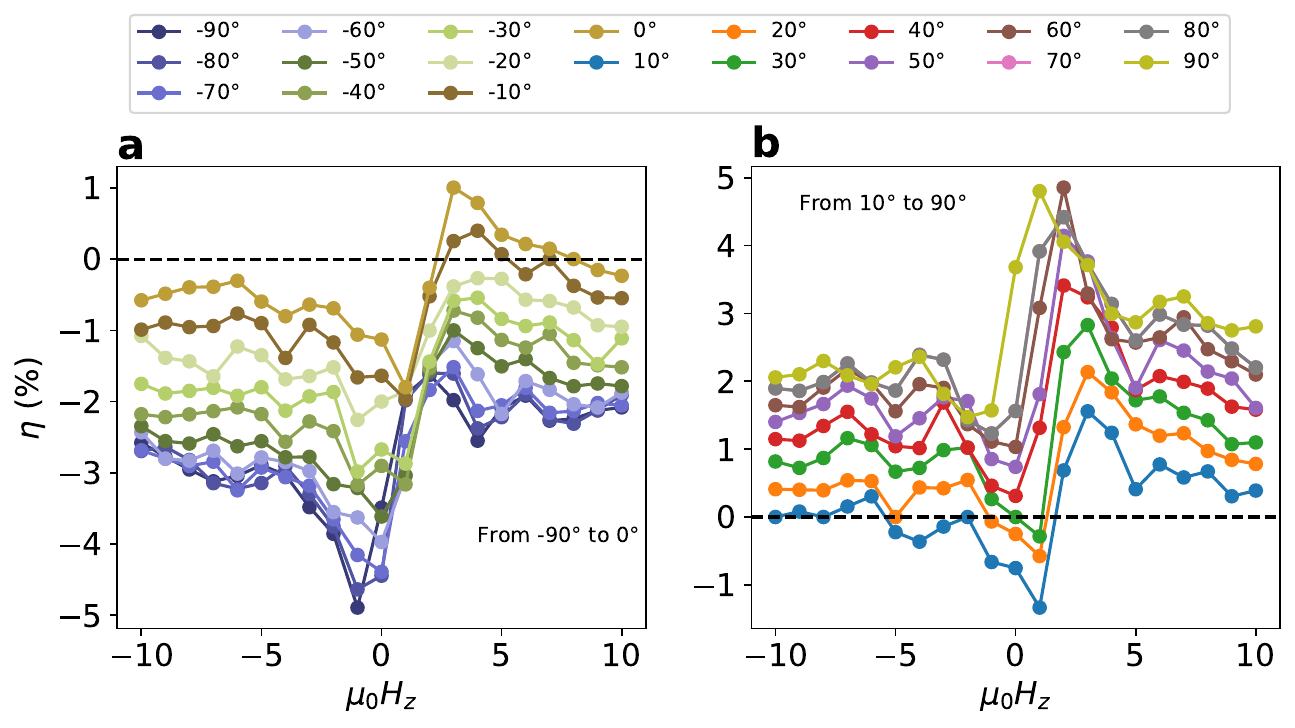}
    \caption{\textbf{SDE efficiency of the Hole device.} SDE efficiency measured under a constant in-plane magnetic field rotating from $-90^\circ$ to $+90^\circ$, alongside a $10$~mT out-of-plane field sweep. \textbf{a}, Measurements for in-plane field angles from $-90^\circ$ to $0^\circ$. \textbf{b}, Measurements for in-plane field angles from $10^\circ$ to $90^\circ$.}
    \label{etain_hole}
\end{figure}

\begin{figure}[H]
    \centering
    \includegraphics[width=\linewidth]{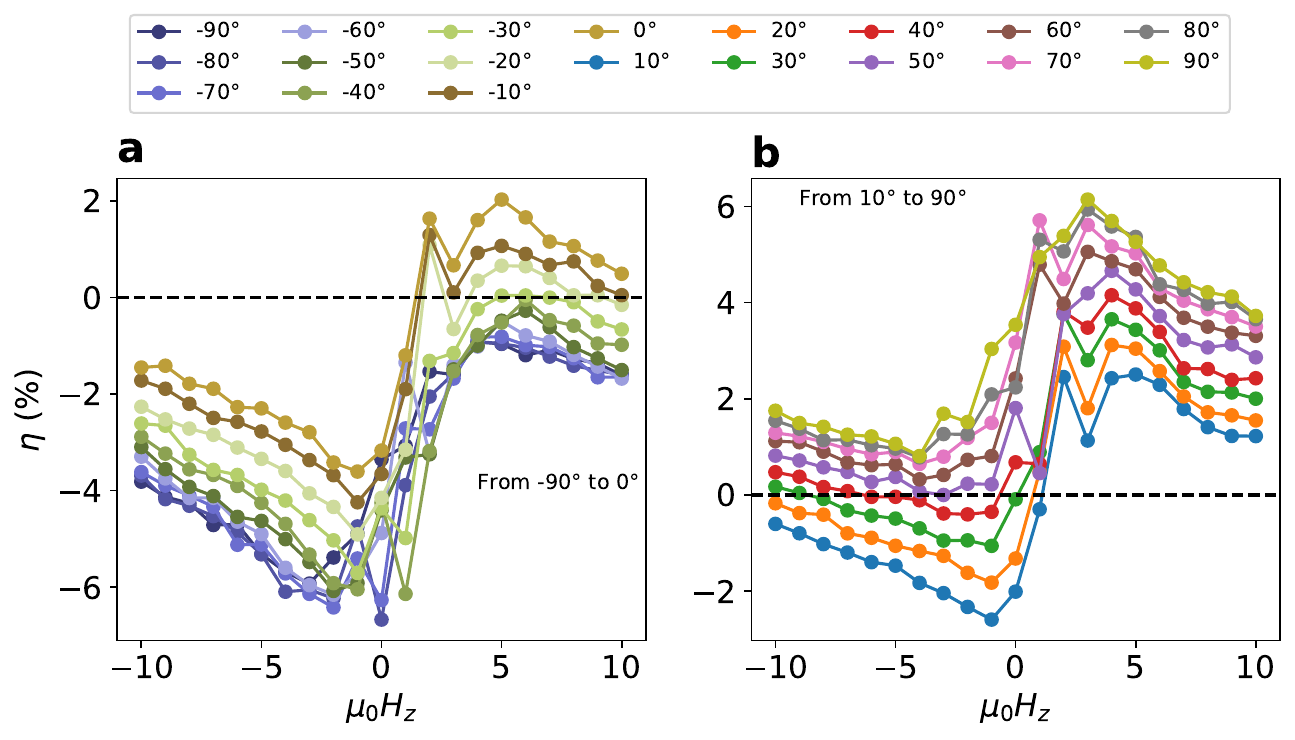}
    \caption{\textbf{SDE efficiency of the Drop device.} SDE efficiency measured under a constant in-plane magnetic field rotating from $-90^\circ$ to $+90^\circ$, alongside a $10$~mT out-of-plane field sweep. \textbf{a}, Measurements for in-plane field angles from $-90^\circ$ to $0^\circ$. \textbf{b}, Measurements for in-plane field angles from $10^\circ$ to $90^\circ$.}
    \label{etain_drop}
\end{figure}

\begin{figure}[H]
    \centering
    \includegraphics[width=\linewidth]{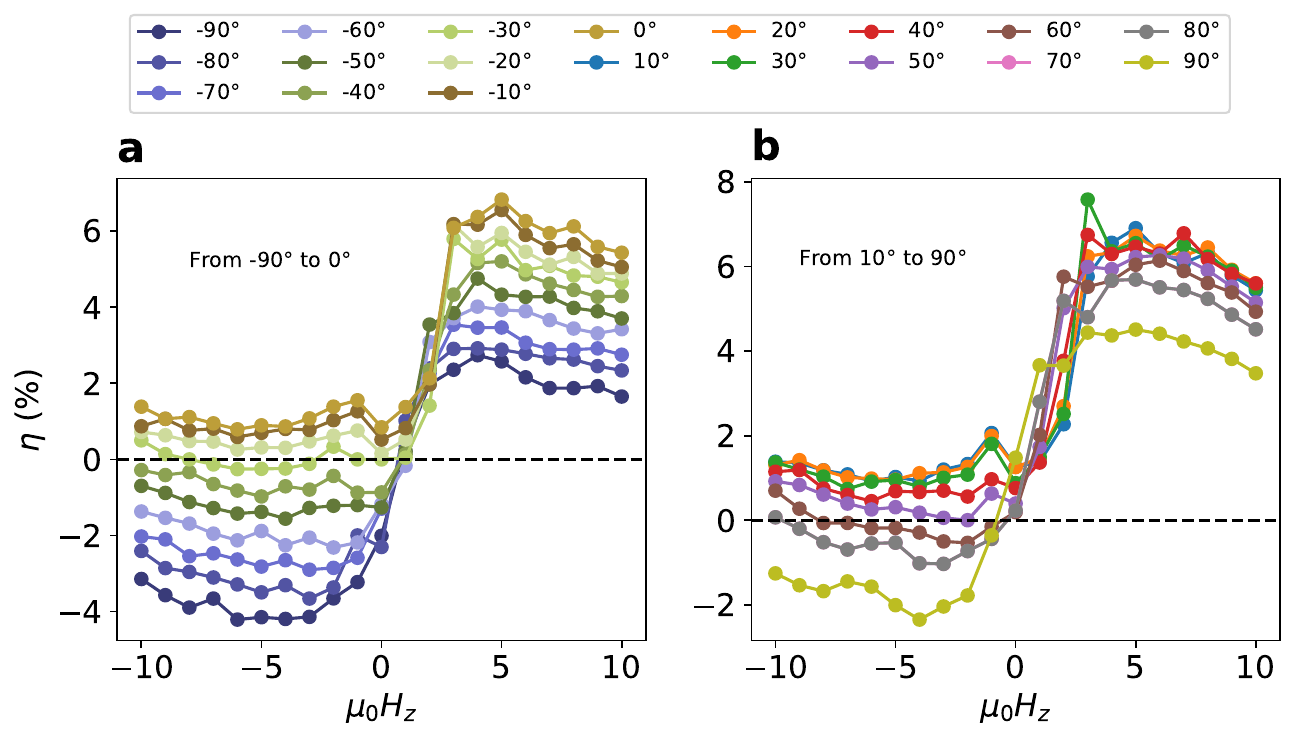}
    \caption{\textbf{SDE efficiency of the Triangle device.} SDE efficiency measured under a constant in-plane magnetic field rotating from $-90^\circ$ to $+90^\circ$, alongside a $10$~mT out-of-plane field sweep. \textbf{a}, Measurements for in-plane field angles from $-90^\circ$ to $0^\circ$. \textbf{b}, Measurements for in-plane field angles from $10^\circ$ to $90^\circ$.}
    \label{etain_tri}
\end{figure}



\end{document}